\documentclass[aps,prl,twocolumn,superscriptaddress,showpacs]{revtex4}

\usepackage{graphicx}

\begin{document}

\title{Eu$_{0.5}$Sr$_{1.5}$MnO$_4$: a three-dimensional XY spin glass}

\author{R. Mathieu\cite{cryo}}
\affiliation{Spin Superstructure Project (ERATO-SSS), JST, AIST Central 4, Tsukuba 305-8562, Japan}

\author{A. Asamitsu}
\affiliation{Spin Superstructure Project (ERATO-SSS), JST, AIST Central 4, Tsukuba 305-8562, Japan}
\affiliation{Cryogenic Research Center (CRC), University of Tokyo, Bunkyo-ku, Tokyo 113-0032, Japan}

\author{Y. Kaneko}
\affiliation{Spin Superstructure Project (ERATO-SSS), JST, AIST Central 4, Tsukuba 305-8562, Japan}

\author{J. P. He}
\affiliation{Spin Superstructure Project (ERATO-SSS), JST, AIST Central 4, Tsukuba 305-8562, Japan}

\author{Y. Tokura}
\affiliation{Spin Superstructure Project (ERATO-SSS), JST, AIST  Central 4, Tsukuba 305-8562, Japan}
\affiliation{Correlated Electron Research Center (CERC), AIST Central 4, Tsukuba 305-8562, Japan}
\affiliation{Department of Applied Physics, University of Tokyo, Tokyo 113-8656, Japan}

\begin{abstract}
The frequency, temperature, and dc-bias dependence of the ac-susceptibility of a high quality single crystal of the Eu$_{0.5}$Sr$_{1.5}$MnO$_4$ layered manganite is investigated. Eu$_{0.5}$Sr$_{1.5}$MnO$_4$ behaves like a XY spin glass with a strong basal anisotropy. Dynamical and static scalings reveal a three-dimensional phase transition near $T_g$ = 18 K, and yield critical exponent values between those of Heisenberg- and Ising-like systems, albeit slightly closer to the Ising case. Interestingly, as in the latter system, the here observed rejuvenation effects are rather weak. The origin and nature of the low temperature XY spin glass state is discussed.
\end{abstract}

\maketitle

The nature of the spin-glass (SG) phase transition has been puzzling experimentalists and theoretists for about 30 years\cite{Binder}. The spin-glass correlation length, as well as many other physical properties diverge at the phase transition temperature $T_g$ with characteristic exponents. The renormalization group theory categorizes phase transitions in different universality classes. Each class is characterized by its own set of critical exponents. For example, the anisotropic Ising SG and the more isotropic vector SG such as XY- and Heisenberg SG belong to different universality classes. For a long time, numerical simulations predicted a phase transition at $T_g$ = 0 K for the XY and Heisenberg systems\cite{NumRes}. More recently, the SG transition was investigated using Ising-like variables, namely the chiralities\cite{Kawamura} defined by the non-collinear spin structure (with left or right handeness). A chiral-glass (CG) phase transition was found at a finite temperature. Depending on the long length-scale coupling\cite{APY} or decoupling\cite{Kawamura} of the spins and chiralities, the CG order may or may not be accompanied by a SG order.\\
\indent Experimentally, a three-dimensional SG phase transition is observed in both Ising and Heisenberg systems at finite $T_g$. For example the (Fe,Mn)TiO$_3$ is a model Ising system\cite{Ito} with a $T_g$ near 22 K over a broad range of (Fe,Mn) compositions, while the so-called canonical SG (dilute magnetic alloys such as Au(Fe), Cu(Mn), or Ag(Mn)\cite{Levy}) are typical Heisenberg-like SG with a $T_g$ depending almost linearly on the amount of the magnetic impurity. In contrast, there are few known true XY SG. The exotic chiral-glass superconductors showing the paramagnetic Meissner effect (PME) are considered as the closest experimental realization of the XY SG\cite{PME2}. This glassy state is unconventional, and often referred to as an orbital-glass state\cite{PME2}, as it involves spontaneous orbital moments rather than spins.\\
\indent The magnetic and electrical properties of the $R_{1-x}A_x$MnO$_3$ manganites ($R$ is a rare earth and $A$ is an alkaline earth element) are essentially controlled by the effective one-electron bandwidth\cite{Tomiokanew}, which characterizes the transfer of the conduction electrons between neighboring Mn sites. Thus, depending on the radii of the $R$ and $A$ cations, the ferromagnetic (FM) metallic phase, or the charge- and orbital-ordered (CO-OO) phase can be stabilized\cite{Orbital,Tomiokanew}. The phase diagram of these perovskites manganites also depends on the degree of quenched disorder\cite{Tomiokanew}. In the layered $R_{1-x}A_{1+x}$MnO$_4$, the long-range CO-OO state near $x$ = 0.5 is gradually suppressed as the ionic radius of $R$ decreases.\cite{Uchida}. La$_{0.5}$Sr$_{1.5}$MnO$_4$ shows the CO-OO transition near $T_{co}$ = 220 K. The CO-OO correlation length decreases already for $R$ = Pr and for  $R$ = Eu, only the nanometer-sized spin and orbital correlation\cite{EBMO} is observed\cite{Uchida}. In the present article, we report the magnetic properties of a single crystal of Eu$_{0.5}$Sr$_{1.5}$MnO$_4$ layered manganite. The system behaves like a three-dimensional XY spin glass. A cusp is observed in the zero-field cooled magnetization only if the probing magnetic field is applied within the $ab$-plane of the crystal. Typical SG features are observed in the time- and temperature dependence of the ac-susceptibility. Dynamical and static scalings are performed, revealing a three-dimensional SG transition at $T_g$ = 18 K. Critical exponents characteristic of the transition are deduced, and compared to those obtained for Ising and Heisenberg systems. The origin of the XY character of the low-temperature SG state is discussed.

High quality single crystals of the $A$-site disordered Eu$_{0.5}$Sr$_{1.5}$MnO$_4$ were grown by the floating zone method. The phase-purity of the crystals was checked by x-ray diffraction. The magnetization $M$ and ac-susceptibility $\chi$($T,\omega=2\pi f$) data were recorded on a MPMSXL SQUID magnetometer equipped with the ultra low-field option (low frequencies) and a PPMS6000 (higher frequencies), after carefully zeroing or compensating the background magnetic fields of the systems. Additional phase corrections were performed for some frequencies. 

\begin{figure}[h]
\includegraphics[width=0.46\textwidth]{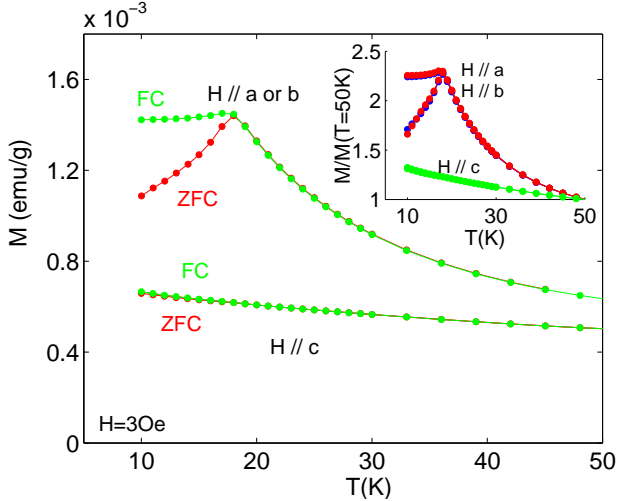}
\caption{(color online) Temperature dependence of the ZFC and FC magnetization recorded in $H$ = 3 Oe applied along the $c$-axis or within the $ab$-plane. The inset shows the normalized magnetization curves for $H$//a, $H$//b, and $H$//c.}
\label{fig-dc}
\end{figure}

\begin{figure}[h]
\includegraphics[width=0.46\textwidth]{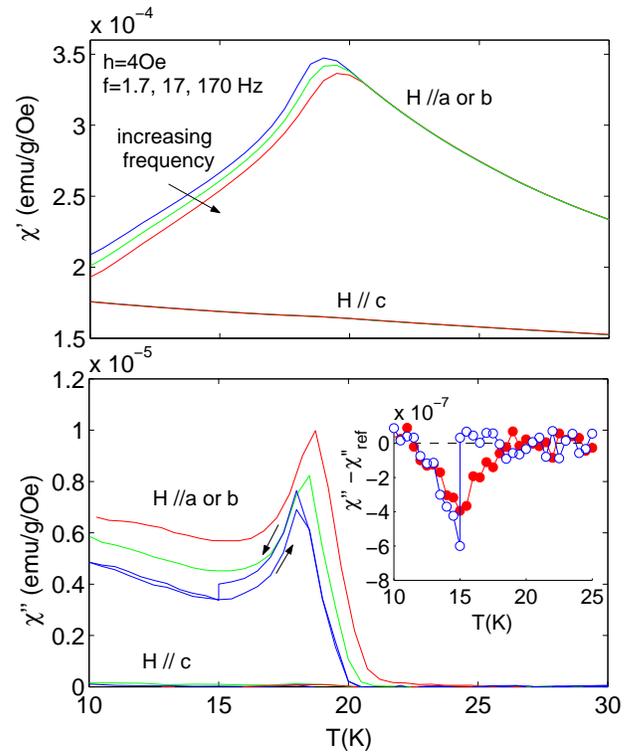}
\caption{(color online) Temperature dependence of the in-phase (upper panel) and out-of-phase (lower panel) components of the ac-susceptibility $\chi'(T,\omega = 2\pi f)$ and  $\chi''(T,\omega)$ for different orientation of the probing field $h$ with $f$ = 1.7, 17, and 170 Hz; $h$ = 4 Oe. The susceptibilities are recorded on heating. In the lower panel, for the lowest frequency, the data collected on a cooling including a 6000 s halt at $T_h$ = 15 K and during the subsequent re-heating is shown to illustrate the aging, memory, and rejuvenation effects (see main text). The inset shows the difference plots of the two curves after subtracting reference curves measured on continuous cooling (open circles) and heating (filled circles).}
\label{fig-ac1}
\end{figure}

Figure~\ref{fig-dc} shows the temperature dependence of the zero-field cooled (ZFC) and field-cooled (FC) magnetization of Eu$_{0.5}$Sr$_{1.5}$MnO$_4$, recorded on heating in a small magnetic field of $H$ = 3 Oe applied in different directions. As seen in the figure and the inset, a cusp is observed below $T$ $\sim$ 20K when $H$ is applied within the $ab$-plane of the Eu$_{0.5}$Sr$_{1.5}$MnO$_4$ crystal. If $H$ is applied along the $c$ direction, no cusp is observed. This XY-like anisotropy is the complete opposite of the Ising SG case, in which a cusp is observed only for $H//c$\cite{Ito}. In addition, as shown in Fig.~\ref{fig-ac1}, the ac-susceptibility $\chi$($T,\omega$) exhibits a fairly large frequency dependence only for $H$ within the $ab$-plane. For $H // c$, no frequency dependence is observed, and the out-of-phase component of the susceptibility $\chi''$($T,\omega$) is almost negligible.

\noindent Eu$_{0.5}$Sr$_{1.5}$MnO$_4$ exhibits typical dynamical SG features, such as aging, memory, and rejuvenations\cite{ghost}. As an illustration, in the lower panel of Fig.~\ref{fig-ac1}, $\chi''$($T,f=1.7 Hz$) is recorded on cooling and heating, after a 6000 s halt at $T_h$ = 15 K (as indicated by the arrows). We observe the decay of $\chi''$ during the halt, reflecting the equilibration of the spin configuration (aging). If we subtract from these data those obtained while continuously changing $T$, we obtain the difference plots shown in the inset of Fig.~\ref{fig-ac1}. One observes that the aging at $T_h$ = 15 K is recovered on re-heating, as the spin configuration established during the equilibration is frozen-in upon cooling below $T_h$ (memory). However, the memory of the equilibration at $T_h$ is observed only in a finite temperature range around $T_h$, defining in the difference plots ``memory dips'' with a finite width. Outside this temperature range, $\chi''$($T,\omega$) recovers its reference level (rejuvenation). Similar glassy features are observed in the three-dimensional cubic perovskite Eu$_{0.5}$Ba$_{0.5}$MnO$_3$\cite{EBMO} case. Eu$_{0.5}$Sr$_{1.5}$MnO$_4$ shows qualitatively the similar weak rejuvenation effect observed in  Eu$_{0.5}$Ba$_{0.5}$MnO$_3$. 
\begin{figure}[h]
 \includegraphics[width=0.46\textwidth]{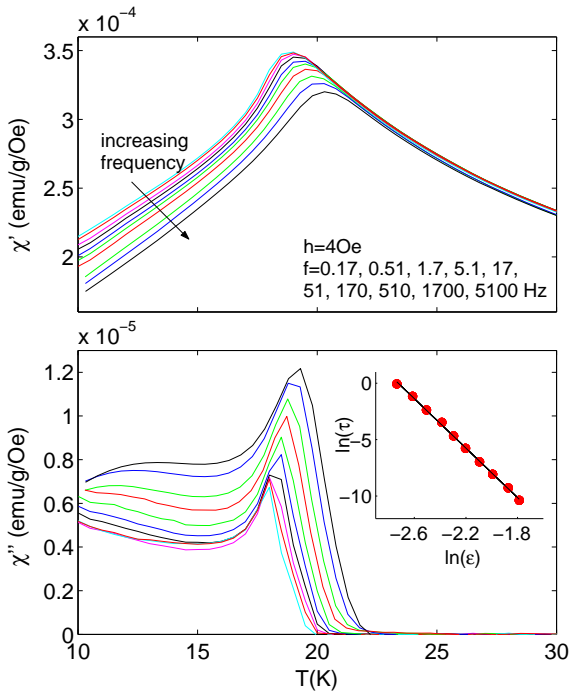}
\caption{(color online) Temperature dependence of the in-phase (upper panel) and out-of-phase (lower panel) components of the ac-susceptibility $\chi'(T,\omega)$ and  $\chi''(T,\omega)$ for $h//a$. The inset shows the dynamical scaling of $\tau$($T_f$) = $t_{obs}$ with the reduced temperature $\epsilon=(T_f(f)-T_g)/T_g$ for  $T_g$ = 18 K, implying $z\nu$ = 10.97 and $\tau_0$ $\sim$ 1.03$\times$10$^{-13}$ s.}
\label{fig-ac2}
\end{figure}
\noindent The memory dips are rather broad, and in continuous measurements (without halts) the heating curves lie slightly below the cooling curves, reflecting the accumulation of the aging (rather than its reinitialization, or rejuvenation). Interestingly, similar accumulative features are observed in the Ising SG\cite{ghost}. In Heisenberg-like SG\cite{ghost}, or in the chiral-glass superconductor\cite{PME3}, rejuvenation effects are much stronger\cite{ghost,PME3}.

The $T$- and $f$-dependence of $\chi''$($T,\omega$) of Eu$_{0.5}$Sr$_{1.5}$MnO$_4$ shown in lower panel of Fig.~\ref{fig-ac2} is analyzed in detail. Each frequency corresponds to an observation time $t_{obs} = 1/\omega$ characteristic of the measurement. One can define from each susceptibility curve a frequency dependent freezing temperature $T_f$($\omega$), below which the longest relaxation time of the system exceeds $t_{obs}$, and the system is out-of-equilibrium. In two dimensions, the dynamic slowing down toward the SG transition is usually described using the generalized Arrhenius law\cite{Huse} for $\tau$($T_f$) = $t_{obs}$: $log(\tau/\tau_0) = 1/T_f^{1+\psi\nu}$, implying $T_g$ = 0 K; $\psi$ and $\nu$ are critical exponents while $\tau_0$ reflects the flipping time of the fluctuating entities. In three dimensions, the SG correlation length $\xi$ diverges at $T_g$ $>$ 0, and $\tau$($T_f$) follows the power law relation\cite{dynscal}: $\tau/\tau_0 = \epsilon^{-z\nu}$ where $\epsilon = (T_f-T_g)/T_g$ is the reduced temperature and $z$ is a critical exponent. Eu$_{0.5}$Sr$_{1.5}$MnO$_4$ has a tetragonal structure, with the typical lattice-parameter ratio $c/a$ $>$ 3 of the layered materials. In these layered manganites, the MnO$_2$ planes ($ab$-planes) are isolated by two blocking $(R$/$Sr)$O layers, so that the CO-OO correlation is limited to the two-dimensional Mn network. One could thus expect a two-dimensional SG state. However, using the two-dimensional expression for the dynamical slowing down, a good scaling is obtained only for unphysical values of $\psi\nu$ and $\tau_0$. $\psi\nu$ here amounts to 6 while $\psi\nu$ $\lesssim$ 1 is expected\cite{cumn,Huse}. As seen in the inset of Fig.~\ref{fig-ac2}, a good scaling is obtained in the three-dimensional case, for $T_g$ = 18 $\pm$ 1 K, $z\nu$ =  11 $\pm 1$.  Interestingly, it was found in SG/metal multilayers, that even a very weak interlayer coupling causes a three-dimensional character of the spin system on experimental time scales\cite{cumn}. This SG state is nearly atomic, as the obtained $\tau_0$ $\sim$ 10$^{-13 \pm 1}$ s is very close to the microscopic spin flip time\cite{EBMO}. The $z\nu$ product obtained for Eu$_{0.5}$Sr$_{1.5}$MnO$_3$ is similar to that obtained in the cubic perovskite case\cite{EBMO}, although $T_g$ is about half\cite{EBMO,Tomiokanew}. The $z\nu$ is also similar to the values obtained for three-dimensional Ising SG (see Table~\ref{tablexp}). 

We now study the non-linear susceptibility of Eu$_{0.5}$Sr$_{1.5}$MnO$_4$ to confirm the phase transition, and estimate the critical exponents of this XY-like system. Higher harmonics than the linear ac-susceptibility are difficult to measure correctly, and thus it is quite difficult to study their divergence at $T_g$\cite{Levy}. One can instead perform a so-called static scaling of the ac-susceptibility, probing the system just above $T_g$ .

\begin{figure}[h]
\includegraphics[width=0.46\textwidth]{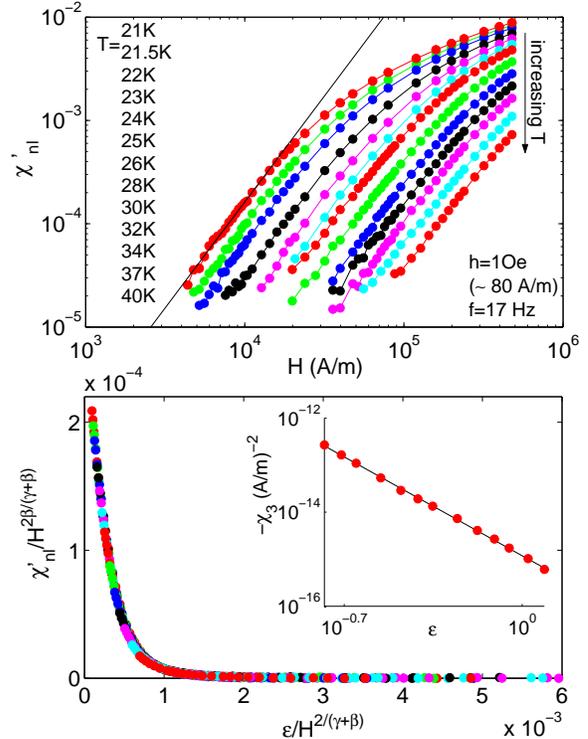}
\caption{(color online) Upper panel: The non-linear susceptibility $\chi'_{nl}$ recorded using the ac field $h$ = 1 Oe as a function of the dc bias field $H$ for different temperatures ($f$=17Hz$,  \chi''_{nl}$ = 0 for all the measured temperatures, just above $T_f(f=17Hz)$). SI units are employed to obtain dimensionless susceptibilities. As indicated by the black line for $T$ = 21 K, the low-field data varies linearly with $H$ (in log-log axis) with a slope of 2.048, indicating that as expected, $\chi'_{nl} \sim  -3 \chi_3$ $H^2$. Lower panel: in the inset, the $\chi_3$ data obtained from the linear fit of the $\chi '_{nl}$ data for all temperatures is scaled with the reduced temperature $\epsilon=(T-T_g)/T_g$, implying $\gamma$ = 3.0 $\pm$ 0.5. The main frame shows the scaling of $\chi'_{nl}/H^{2\beta/(\gamma + \beta)}$ with $\epsilon/H ^{2/(\gamma + \beta)}$, yielding $\beta$ = 0.5 $\pm$ 0.1.}
\label{fig-stat} 
\end{figure}

The magnetization $M$ in a spin glass can be expressed in odd powers of the magnetizing field $H$ as $M=\chi_0H+\chi_3H^3+\chi_5H^5+...$. $\chi_1$ is the linear susceptibility and the non-linear susceptibility $\chi_{nl}=\chi_1-M/H$ contains the higher-order terms. Unlike $\chi_1$, $\chi_3$ and all the high-order terms diverge close to $T_g$\cite{Susuki}. If we probe the system using a small ac field $h {\rm sin}(\omega t)$ as a function of a superposed dc bias field $H>>h$, the in-phase component of the ac-susceptibility $\chi'(T,\omega)$ can be written\cite{Levy} as $\chi'(T,\omega)=\chi_1+3\chi_3H^2+5\chi_5H^4+...$ so that $\chi'_{nl}(T,\omega)=-3\chi_3H^2-5\chi_5H^4-...$, and thus for the smaller bias fields, $\chi'_{nl}(T,\omega) \sim -3\chi_3H^2$. The upper panel of Fig.~\ref{fig-stat} shows the non-linear susceptibility $\chi'_{nl}$ extracted as explained above as a function of the dc bias field $H$ for different temperatures. The linear fit (in log-log axes) of the low-field data for all $T$ yields $\chi_3(T)$.  $\chi_3(T)$ is expected to diverge\cite{Susuki} at $T_g$ as $\chi_3 \propto \epsilon^{-\gamma}$, where $\epsilon = (T-T_g)/T_g$ is the reduced temperature and $\gamma$ is a critical exponent. This divergence is observed, as shown in the inset of the lower panel of Fig.~\ref{fig-stat}, yielding $T_g$ = 18 K and $\gamma$ = 3 $\pm$ 0.5.  In contrary to pure or dilute ferromagnets or antiferromagnets, it is not possible to probe the equilibrium properties very close to $T_g$ in a SG, since dynamical effects already contribute to the measured susceptibility\cite{Gurkan,Mattsson} (deviations from pure $H^2$ behavior are observed closer to $T_g$). In our scaling analysis, the smallest reduced temperature corresponds to a relaxation time $\tau=\tau_0\epsilon^{-z\nu}$ $\sim$ 3.7 $\times$ 10$^{-5}$ s, which is several orders of magnitude lower than the timescale of our measurement (1/$\omega$ $\sim$ 0.01s)\cite{note}. While it could be argued that the region of critical behavior is much extended in reduced temperatures in SG than in pure magnets, it is not possible to use larger reduced temperatures either\cite{Tomas}.

A scaling relation of the non-linear susceptibility in the critical region was proposed\cite{Susuki}, and modified\cite{Gesch} as: $\chi_{nl} = H^{2\beta/(\gamma + \beta)} G[\epsilon/H ^{2/(\gamma + \beta)}]$, where $\beta$ is another critical exponent. We obtained a good collapse of the all the data for $\beta$ = 0.5 $\pm$ 0.1, as shown in Fig.~\ref{fig-stat}. The critical exponents describing the second-order phase transition are related by various scaling laws\cite{Kaldanoff}, and can thus be estimated from $\gamma$ and $\beta$. For example, the specific-heat exponent $\alpha$ can be calculated using the relation $\alpha + 2\beta + \gamma = 2 $. Similarly, $\delta$, $\nu$ (and thus $z$ using the results of the dynamical scaling), and $\eta$ which governs the spatial correlation-function near $T_g$, can be estimated. The values obtained for Eu$_{0.5}$Sr$_{1.5}$MnO$_4$ are indicated in Table~\ref{tablexp}. As seen in the table, the values of the critical exponents of Eu$_{0.5}$Sr$_{1.5}$MnO$_4$ are quite close to those of the Ising SG, but essentially lie between those obtained for the Heisenberg and Ising SG.

\begin{table}
\caption{Critical exponents for the three-dimensional Heisenberg-like Ag(Mn) and Ising-like (Fe,Mn)TiO$_3$ spin glasses (from Refs. \onlinecite{Levy}, \onlinecite{Gurkan}, \onlinecite{FisherHz}), and Eu$_{0.5}$Sr$_{1.5}$MnO$_4$. The uncertainty on the exponents is given in parenthesis.}
\label{tablexp}
\begin{tabular}{cccc}
&Heisenberg&Eu$_{0.5}$Sr$_{1.5}$MnO$_4$&Ising\\
\colrule
$\gamma$&2.3&3 (0.5)&4\\
$\beta$&0.9&0.5 (0.1)&0.54\\
$\alpha$&-2.1&-2 (0.5)&-3\\
$\nu$&1.3&1.3 (0.2)&1.7\\
$\delta$&3.3&7 (1.5)&8.4\\
$\eta$&0.4&-0.25 (0.07)&-0.35\\
\colrule
$z\nu$&6-8&11 (1)&10-12\\
\end{tabular}
\end{table}

The SG state of Eu$_{0.5}$Sr$_{1.5}$MnO$_4$ originates from the fragmentation of the CO-OO state down to the nanometer scale, which causes the mixture of antiferromagnetic and FM bonds on similar length scales\cite{EBMO}. The anisotropy of the SG states gives some insight on the low temperature orbital state. One may for example speculate that the short-ranged CO-OO state of Eu$_{0.5}$Sr$_{1.5}$MnO$_4$ mainly includes $3x^2-r^2/3y^2-r^2$ orbitals, favoring inplane magnetic moments\cite{Orbital,Uchida}. Thus the fluctuating entities may comprise more than one single atomic spin. However the total number of spins must be small as the $\tau_0$ obtained from the dynamical scaling is atomic-like ($\sim$ 10$^{-13}$ s). In superspin glasses, weak rejuvenation effects are observed as well\cite{Petra}. However in those interacting magnetic nanoparticle systems, the fluctuating entities are very large, $\sim$ 10$^3$ coherent atomic spins, with a flipping time $\tau_0$ $\sim$ 10$^{-4}$-10$^{-8}$ s.

In summary, we have found that Eu$_{0.5}$Sr$_{1.5}$MnO$_4$ behaves like a XY spin glass system, with a finite phase transition at $T_g$ = 18 K. We believe that this anisotropic spin-glass state originates from the two-dimensional Mn network in the layered manganites, and the associated orbital ordering. The critical exponents associated with the phase transition are obtained from the dynamical and static scalings of the ac-susceptibility. These exponents lie, more or less, between those of the Heisenberg- and Ising SG systems, albeit the dynamical glassy features seem closer to those observed in Ising systems than in the Heisenberg ones.

We thank Prof. P. Nordblad for his expert advices.


\begin{thebibliography}{10}
\bibitem{note}
As the relaxation time scales with the correlation length as $\tau \propto \xi^z$, it can also be deduced that the length scales probed in our measurements range from 2 to 10 times the fluctuating length scales.
\end{thebibliography}
\end{document}